\documentclass[preprint,showpacs,nofootinbib]{revtex4}
\usepackage{amsmath}
\usepackage{graphicx}
\usepackage{physics}
\newcommand{\diff}{\mathop{}\!\mathrm{d}}
\newcommand{\nicevec}[1]{\Vec{\boldsymbol{#1}}}

\newcommand{\traceless}[1]{\overset{\circ}{\overline{#1}}}

\begin{document}

\title{Shear viscosity of a massless quark-gluon gas in chemical
  equilibrium in terms of all $2\leftrightarrow 2$ cross sections} 

\author{Okey Ohanaka}
\affiliation{Department of Physics, East Carolina University,
  Greenville, NC 27858, USA} 
\author{Zi-Wei Lin}
 \email{linz@ecu.edu}
\affiliation{Department of Physics, East Carolina University,
  Greenville, NC 27858, USA} 

\date{\today}

\begin{abstract}
The analytical expressions of the shear viscosity of both one and two 
particle species with Boltzmann statistics and $2 \rightarrow 2$ elastic
scatterings are known from the Chapman-Enskog method and have been
shown to be quite accurate. The expression for a multi-species 
hadronic gas under $2 \rightarrow 2$ elastic scatterings is also
known. Here we use the Chapman-Enskog method to derive the 
explicit expression of shear viscosity of a massless quark-gluon gas
of $N_f$ quark flavors in chemical equilibrium subjected to all $2
\rightarrow 2$ parton scatterings including for the first time
inelastic scatterings.  We then verify the expression in a general
single-species limit, where the shear viscosity of the quark-gluon gas
should reduce to the result for a single particle species. 
In addition, we show the explicit analytical result in terms of the
seven independent cross sections for the special case of isotropic and
energy-independent cross sections. 
The analytical expressions derived here can be useful for determining 
the shear viscosity of parton transport models with any $2 \rightarrow
2$ scattering cross sections. They can also be coupled with finite
temperature QCD cross sections to help study the shear viscosity of the
quark gluon plasma.
\end{abstract}

\maketitle

\section{Introduction}

The quark-gluon plasma (QGP) is a state of matter achieved when
conditions are sufficient for quarks and gluons to become deconfined
from hadrons. The QGP is believed to have been created in heavy ion
collisions at the Large Hadron Collider (LHC) and the Relativistic
Heavy Ion Collider (RHIC)~\cite{arsene2005first}. 
The QGP is also theorized to be the state of matter of the universe
less than $10^{-5}$ seconds after the Big
Bang~\cite{Snellings:2011sz}, so understanding its properties will
enable a better understanding of the universe's evolution. 
Observables from heavy ion collisions including the anisotropic flows 
indicate that the QGP is a near perfect fluid with a very small shear
viscosity to entropy density ratio $\eta/s$.
This ratio has been conjectured to have a lower bound of
$1/(4\pi)$~\cite{Kovtun:2004de}. Extractions of the shear viscosity 
from comparisons between theoretical models and the experimental 
data~\cite{Ferini:2008he,Bernhard:2015hxa,Bernhard:2019bmu,Parkkila:2021tqq}
have shown that the QGP $\eta/s$ within a few times of 
the QCD phase transition temperature ($T_c \simeq 156~\mathrm{MeV}$)
is not too far from the conjectured lower bound, although the
uncertainty is large and its temperature dependence cannot be yet
determined well. 

While the QGP shear viscosity including its temperature dependence is
an input function in hydrodynamical models of heavy ion collisions,
it is not an input in transport models 
\cite{Zhang:1997ej,Lin:2004en,Xu:2007ns,Kurkela:2022qhn,Parisi:2025gwq}. 
Instead, it is implicitly determined by the interactions among quarks
and gluons. Therefore, the ability to compute the shear viscosity within
the quasi-particle picture is important for the development and 
application of the transport model
approach~\cite{Zhang:1997ej,Lin:2004en,Xu:2007ns,MacKay:2022uxo,Kurkela:2022qhn,Parisi:2025gwq}
and the QCD effective kinetic theory
approach~\cite{Kurkela:2018vqr}. These two approaches are  both well
suited for the description of non-equilibrium dynamics, 
which is expected to be important for small collision systems as well
as early times or heavy flavor observables in large collision
systems. An example is the parton escape mechanism for anisotropic
flows~\cite{He:2015hfa,Lin:2015ucn}, which can efficiently
convert spatial anisotropies into momentum anisotropies via a small 
number of scatterings far from equilibrium. 
An extensive computation of the shear viscosity of the QGP at high 
temperatures has been performed in the AMY framework developed by
Arnold, Moore and Yafee~\cite{Arnold:2000dr,Arnold:2003zc},  
where both $2\leftrightarrow2$ and $1\leftrightarrow 2$ interactions
have been taken into account and the $2\leftrightarrow2$ perturbative
QCD (pQCD) cross sections are screened with thermal self energies. 
When one extrapolates the AMY results to finite temperatures below  
several $T_c$, the $\eta/s$ values are much higher than $1/(4\pi)$ or  
the current $\eta/s$ range extracted from the experimental
data. On the other hand, next-to-leading order corrections have been
found to be large and significantly decrease the $\eta/s$ value at
physically relevant temperatures~\cite{Ghiglieri:2018dib}. 

Regarding analytical relations between the shear viscosity and
arbitrary scattering cross sections, there are several methods to
calculate the shear viscosity for a single particle species with
Boltzmann statistics and $2 \rightarrow 2$ elastic
scatterings~\cite{Wiranata:2012br,Plumari:2012ep}, 
including the relaxation time approximation 
and the Chapman-Enskog (CE) method. For a massless 
parton gas at temperature $T$ that scatters with an isotropic and
energy-independent (elastic) cross section $\sigma$, different methods 
give essentially the same result $\eta_1^{\rm iso}=6T/(5\sigma)$. For
anisotropic scatterings, however, only the Chapman-Enskog method is
accurate enough based on the comparisons with numerical results
obtained using the Green-Kubo
relation~\cite{Plumari:2012ep,MacKay:2022uxo}. The calculation of
shear viscosity for a hadronic resonance gas with elastic $2
\rightarrow 2$ scatterings~\cite{Wiranata:2013oaa} provides the
analytical result for a multi-species system, where the explicit
$\eta$ expression for a binary mixture was
given~\cite{Wiranata:2013oaa} and has been recently shown to agree
well with the Green-Kubo numerical results~\cite{Parisi:2025gwq}. Note
that another study~\cite{Moroz:2014gda} addressed the special case of
isotropic and energy-independent elastic $2 \rightarrow 2$ scatterings
for a multi-species system. 

The goal of this work is to derive an explicit analytical relationship
between the shear viscosity of a massless quark-gluon gas in chemical 
equilibrium and all the $2 \leftrightarrow 2$ differential cross
sections including the inelastic channels. 
We will use the Boltzmann statistics for the computation. The
paper is organized as follows. After the introduction, we will review
in Sec.~\ref{method} the method used for this work. Then we will
consider all parton $2 \leftrightarrow 2$ scatterings and derive the
shear viscosity expression in Sec.~\ref{results}. 
In Sec.~\ref{limit} we verify the general $\eta$ expression
in the single-species limit, where the shear viscosity should reduce to 
the single-species result, and we also give a more explicit $\eta$
expression in the special case of isotropic and energy-independent
cross sections. After some discussions in Sec.~\ref{discussion}, 
we summarize in Sec.~\ref{conclusion}.

\section{The Chapman-Enskog Method for shear viscosity}
\label{method}

In this section, we closely follow the book by de Groot, van Leeuwen
and van  Weert~\cite{deGroot1980} while showing the key steps in the
derivations. We begin by defining the following variables:
\begin{equation}
n \equiv\sum_i n_i, ~x_i \equiv \frac{n_i}{n}, ~z_i \equiv  \frac{m_i}{T}, 
~\pi_i^{\mu} \equiv \frac{p_i^{\mu}}{T}, 
~\tau_i \equiv \frac{p_i^{\mu} U_{\mu}}{T}. 
\end{equation}
In the above, $n_i$ is the number density for particle species
$i$ (from 1 to $N$), $n$ is the total number density of all $N$ species,
$m_i$ is the particle mass that we shall take to zero for this study,
$p_i^{\mu}$ is the particle 4-momentum, and $U^{\mu}$ is the
4-velocity of the gas. We also define the projection tensor 
\begin{equation}
\Delta^{\mu\nu} \equiv g^{\mu\nu} - U^{\mu} U^{\nu},
\label{deltamunu}
\end{equation}
which generates from any tensor $a^{\mu\nu}$ the following
symmetric and traceless tensor: 
\begin{equation}
a\traceless{{}^{\mu\nu}} \equiv
\tfrac{1}{2}\left({\Delta_{\alpha}^{\mu}} {\Delta_{\beta}^{\nu}} 
+ {\Delta_{\beta}^{\mu}} {\Delta_{\alpha}^{\nu}}\right)
a^{\alpha\beta} - \tfrac{1}{3} \Delta^{\mu\nu} \Delta_{\alpha\beta} 
\; a^{\alpha\beta}.  
\end{equation}
Similarly, one can generate from a vector product $b^{\mu}b^{\nu}$ 
a symmetric and traceless $b\traceless{{}^\mu b^\nu}$. 
We will also make use of the following inner
product and collision bracket~\cite{deGroot1980}: 
\begin{equation}
\left\langle F_i ,G_i\right\rangle_i \equiv \frac{T}{n_i}
\int \frac{\diff^3 \nicevec{p}_i}{p_i^0} F_i(p_i) G_i(p_i) f_i(p_i)
=\frac{1}{8\pi T^2} \int \frac{\diff^3
\nicevec{p}_i}{p_i^0} F_i(p_i) G_i(p_i) e^{-\tau_i}  , 
\end{equation}
\begin{equation}
[F,G]_{ij\rightarrow kl} = \frac{1}{128\pi^2 T^6} \int \frac{\diff^3
  \nicevec{p}_i}{p_i^0} \frac{\diff^3 \nicevec{p}_j}{p_j^0} \frac{\diff^3
\nicevec{p}_k}{p_k^0} \frac{\diff^3 \nicevec{p}_l}{p_l^0} \,
e^{-(\tau_i + \tau_j)} F G \, W_{ij\rightarrow kl}. 
\label{fgbracket}
\end{equation}
In the above, $n_i=d_i z_i^2K_2(z_i)T^3/(2\pi^3)$
is the number density of parton species $i$ with mass $z_iT$, 
and we have used the relation $z_i^2K_2(z_i)=2$ as $z_i
\rightarrow 0$ to reduce it for the massless case, where $K_n(x)$
represents the modified Bessel of the second kind. 
The term $f_i(p_i) \equiv dN_i/(d^3xd^3p) =d_i \,
e^{-\tau_i}/(8\pi^3)$ is the equilibrium distribution for Boltzmann
statistics with $d_i$ being the degeneracy factor, and
$W_{ij\rightarrow kl}$ is the collision kernel defined as
\begin{equation}
W_{ij\rightarrow kl} \equiv s \; \delta^{(4)}(p_i + p_j - p_k - p_l)
\frac{\diff\sigma}{\diff\Omega}^{ij\rightarrow kl}, 
\end{equation}
where $s$ is the Mandelstam variable (from now on) and
$\diff\sigma/\diff\Omega$ is the differential cross section for the
scattering $i+j \rightarrow k+l$. 

Referring to previous
works~\cite{deGroot1980,van1973relativistic,van1975relativistic}, 
the first Chapman-Enskog approximation allows us to write the shear
viscosity as 
\begin{equation}
\label{viscosityInnerProduct}
\eta = \frac{T}{10} \sum_i x_i \big\langle \pi\traceless{{}_i^\mu
\pi_i^\nu} \, , \; C_i(\tau_i) \pi\traceless{{}_{i\mu} \pi_{i\nu}} \big\rangle_i. 
\end{equation}
The $C_i(\tau_i)$ above is the collision term specified by
\cite{deGroot1980} 
\begin{equation}
\label{eqnCitaui}
\sum_{j=1}^N x_j R_{ij}[C_i(\tau_i)\pi\traceless{{}_i^\mu \pi_i^\nu}]
=\pi\traceless{{}_i^\mu \pi_i^\nu}, 
\end{equation}
where
\begin{equation}
R_{ij}[\phi_i]=\frac{1}{16\pi T^4} \sum_{k,l} 
\int \frac{\diff^3 \nicevec{p}_j}{p_j^0} \frac{\diff^3
\nicevec{p}_k}{p_k^0} \frac{\diff^3 \nicevec{p}_l}{p_l^0}
e^{-\tau_j} (\phi_i + \phi_j - \phi_k - \phi_l) W_{ij\rightarrow kl}. 
\end{equation}
Applying the above to Eq.\eqref{viscosityInnerProduct}, we get 
\begin{equation}
\begin{aligned}
\eta &= \frac{T}{10} \sum_{i,j} x_i x_j \big\langle
R_{ij}[C_i(\tau_i)\pi\traceless{{}_i^\mu \pi_i^\nu}] \, , \;
C_i(\tau_i) \pi\traceless{{}_{i\mu} \pi_{i\nu}} \big\rangle_i \\ 
&= \sum_{i,j,k,l}  \frac{x_i x_j}{1280 \pi^2 T^5}\int \frac{\diff^3
\nicevec{p}_i}{p_i^0} \frac{\diff^3 \nicevec{p}_j}{p_j^0}
\frac{\diff^3 \nicevec{p}_k}{p_k^0} \frac{\diff^3
\nicevec{p}_l}{p_l^0} \, e^{-(\tau_i + \tau_j)} W_{ij\rightarrow kl} \\ 
& \times \left [
  C_i(\tau_i)\pi\traceless{{}_i^\mu \pi_i^\nu} +
  C_j(\tau_j)\pi\traceless{{}_j^\mu \pi_j^\nu} - 
C_k(\tau_k)\pi\traceless{{}_k^\mu \pi_k^\nu} -
C_l(\tau_l)\pi\traceless{{}_l^\mu \pi_l^\nu} \right ]
C_i(\tau_i) \pi\traceless{{}_{i\mu} \pi_{i\nu}}\, . 
\end{aligned}
\end{equation} 

Approximating $C_i(\tau_i)$ as a trial function of a finite power
series up to power $P$~\cite{deGroot1980} 
\begin{equation}
\label{eqnCiexpansion}
C_i(\tau_i) \approx C_i^{(P)}(\tau_i) = \sum_{r=0}^{P}
C_{i,r}^{(P)} \tau_i^r,  
\end{equation}
and rearranging the indices to extract the constant term
$C_{i,r}^{(P)}C_{j,s}^{(P)}$ (for example, by replacing $C_{i,s}^{(P)}$
with $\sum_m \delta_{im} C_{m,s}^{(P)}$ for the term $C_{i,r}^{(P)}
C_{i,s}^{(P)}$), we can write the shear viscosity as  
\begin{equation}
\eta = \frac{T}{10} \sum_{r,s} \sum_{i,j} C_{ij}^{rs} C_{i,r}^{(P)} C_{j,s}^{(P)},
\end{equation}
where the matrix elements are defined in terms of the collision brackets as
\begin{equation}
\begin{aligned}
\label{eqnCijrs}
C_{ij}^{rs} \equiv 
&\; \delta_{ij} x_i \sum_{k,l,m} x_m [\tau_i^r
\pi\traceless{_i^{\mu} \pi_i^{\nu}} \, , \; \tau_i^s
\pi\traceless{_{i\mu} \pi_{i\nu}} ]_{im\rightarrow kl} 
+ x_i x_j \sum_{k,l} [\tau_i^r
\pi\traceless{_i^{\mu} \pi_i^{\nu}} \, , \; \tau_j^s
\pi\traceless{_{j\mu} \pi_{j\nu}} ]_{ij\rightarrow kl} \\ 
&-x_j  \sum_{k,m} x_m [\tau_i^r
\pi\traceless{_i^{\mu} \pi_i^{\nu}} \, , \; \tau_j^s
\pi\traceless{_{j\mu} \pi_{j\nu}} ]_{jm\rightarrow ik}  
-x_j  \sum_{k,m} x_m  [\tau_i^r \pi\traceless{_i^{\mu}
\pi_i^{\nu}} \, , \; \tau_j^s \pi\traceless{_{j\mu} \pi_{j\nu}}
]_{jm\rightarrow ki}.
\end{aligned}
\end{equation}
Note that the above is somewhat different~\cite{Ohanaka:thesis}
from equation (VI.3.100) of the book~\cite{deGroot1980} in that the
last two terms show up in the book as $-2x_j  \sum_{k,m} x_m
[\tau_i^r \pi\traceless{_i^{\mu} \pi_i^{\nu}}, \tau_j^s
\pi\traceless{_{j\mu} \pi_{j\nu}} ]_{jm\rightarrow ik}$; they would be 
the same if $W_{ij\rightarrow   kl}=W_{ij\rightarrow lk}$, but this is not 
the case for certain pQCD $2 \rightarrow 2$ scatterings including $g
q  \rightarrow g q$ and $q \bar q \rightarrow q \bar q$.  

Equations\eqref{viscosityInnerProduct} and \eqref{eqnCiexpansion}
also allow us to write the shear viscosity as 
\begin{equation}
\label{viscosityCir}
\eta = \frac{T}{10} \sum_{i=1}^N \sum_{r=0}^P x_i \gamma_{i,r}
C_{i,r}^{(P)},  
\end{equation}
where $\gamma_{i,r} \equiv \big\langle\tau_i^r
\pi\traceless{_i^{\mu} \pi_i^{\nu}} , \pi\traceless{_{i\mu}
\pi_{i\nu}} \big\rangle_i$. 
For massless partons, $\gamma_{i,r} \!=\! (r+5)!/3$ 
is independent of the particle species $i$, e.g., $\gamma_{i,0} \equiv
\gamma_0 = 40$ for $r=0$. 
Next we apply the inner product 
$\langle F, x_i \tau_i^r \pi\traceless{_{i\mu} \pi_{i\nu}}
\rangle_i$ to Eq.\eqref{eqnCitaui}, where $F$ represents either side
of the equation, and we obtain~\cite{deGroot1980} (see Appendix A for
details)  
\begin{equation}
\label{eqnCir}
    \sum_{j=1}^N \sum_{s=0}^P C_{ij}^{rs} C_{j,s}^{(P)} = x_i
    \gamma_{i,r}.
\end{equation}

In this study, we only derive the first-order ($P=0$) 
Chapman-Enskog expression, i.e., we assume that the collision
coefficient $C_i(\tau_i)$ is constant with respect to energy. Note
that it has been shown for a single particle species that adding
higher-order terms only changes the result by several
percent~\cite{Wiranata:2012br}. For $P=0$, Eq.\eqref{eqnCir} gives 
\begin{equation}
\nicevec{c} = {\mathbf C}^{-1} \nicevec{\gamma},
\end{equation}
where, for a system of massless gluons and $N_f$ flavors of
massless quarks, we write
\begin{align}
\nicevec{c} \equiv 
\begin{pmatrix}
C_{g,0}^{(0)} \\   C_{q_1,0}^{(0)}  \\  C_{\bar{q}_1,0}^{(0)}  \\
\vdots \\ C_{q_{N_f},0}^{(0)}  \\ C_{\bar{q}_{N_f},0}^{(0)} 
\end{pmatrix} \! \!, 
\nicevec{\gamma} \equiv \gamma_0 
\begin{pmatrix}
x_g \\ x_{q_1} \\ x_{\bar{q}_1} \\ \vdots \\ x_{q_{N_f}} \\
x_{\bar{q}_{N_f}} 
\end{pmatrix} \! \!,
{\mathbf C} \equiv 
\begin{pmatrix}
C_{gg}^{00} & C_{g q_1}^{00} & C_{g \bar{q}_1}^{00} & \cdots & C_{g
  q_{N_f}}^{00} & C_{g \bar{q}_{N_f}}^{00} \\  
C_{q_1g}^{00} & C_{q_1 q_1}^{00} & C_{q_1 \bar{q}_1}^{00} & \cdots &  
C_{q_1 q_{N_f}}^{00} & C_{q_1 \bar{q}_{N_f}}^{00} \\ 
C_{\bar{q}_1 g}^{00} & C_{\bar{q}_1 q_1}^{00} & C_{\bar{q}_1
  \bar{q}_1}^{00} & \cdots  & C_{\bar{q}_1 q_{N_f}}^{00} &
C_{\bar{q}_1 \bar{q}_{N_f}}^{00} \\  
\vdots & \vdots & \vdots & \ddots & \vdots & \vdots \\ 
C_{q_{N_f} g}^{00} & C_{q_{N_f} q_1}^ {00} & C_{q_{N_f} \bar{q}_1}^{00} & \cdots &  C_{q_{N_f} q_{N_f}}^{00} & C_{q_{N_f}
  \bar{q}_{N_f}}^{00} \\  
C_{\bar{q}_{N_f} g}^{00} & C_{\bar{q}_{N_f} q_1 }^{00} &
C_{\bar{q}_{N_f} \bar{q}_1}^{00} & \cdots  & C_{\bar{q}_{N_f} q_{N_f} }^{00}  
& C_{\bar{q}_{N_f} \bar{q}_{N_f}}^{00}
\end{pmatrix} \! \! .
\end{align}
Equation\eqref{viscosityCir} then gives the shear viscosity for
$P=0$ as 
\begin{equation}
\eta = \frac{T}{10} \nicevec{\gamma}^{\rm T} \cdot \nicevec{c} = 
\frac{T}{10} \nicevec{\gamma}^{\rm T} {\mathbf C}^{-1}
\nicevec{\gamma}.
\label{etaVector}
\end{equation}

\section{Explicit results for a massless quark-gluon gas}
\label{results}

We now consider a system of massless gluons and $N_f$ flavors of
quarks in chemical equilibrium, where we write $x_{q_i} \equiv x_q$,
$x_{\bar {q_i}} \equiv x_{\bar q}$ for $i \in [1,N_f]$ with 
$x_q=x_{\bar q}$.  For convenience, we write the matrix elements of
${\mathbf C}$ in the following forms~\cite{Wiranata:2013oaa}: 
\begin{align}
C_{ii}^{00}= x_i^2\,\Tilde{c}_{0}[i] + x_i \sum_{j \neq i}
x_j \, \Tilde{c}_{1}[ij] + \Tilde{\omega}[ii], \; \;
C_{ij}^{00} &= x_i x_j \, \Tilde{c}_{2}[ij] + \Tilde{\omega}[ij] 
\; \; ({\rm for}~i\neq j).
\end{align}

Following the steps shown in Appendix B, for massless partons we obtain
the elastic terms as follows (with $v=\sqrt{s}/T$): 
\begin{equation}
\Tilde{c}_{0}[i] = \frac{1}{384} \int_0^{\infty}
\diff v\, v^6 \left [ (3v^2+4)K_3(v) - 6v\,
      K_2(v) \right ] \sigma_{\rm tr}^{ii\rightarrow ii}(v), 
\label{c0g}
\end{equation}
\begin{multline}
    \Tilde{c}_{1}[ij] = - \frac{1}{192} \int_0^{\infty}
\diff v\, v^2 \, \int_{-v^2 T^2}^{0} \diff t\, \tfrac{t}{T^2} 
    \bigg\{\left[3v^2\left(\tfrac{t}{T^2} + v^2\right) +
      4\left(\tfrac{t}{T^2} + 6v^2\right)\right] K_3(v) \\ 
    - 2v\left(\tfrac{3t}{T^2} - 2v^2\right) K_2(v)\bigg\}
    \frac{\diff\sigma}{\diff t}^{ij\rightarrow ij}, 
\end{multline}
\begin{multline}
    \Tilde{c}_{2}[ij] = -\frac{1}{192} \int_0^{\infty}
\diff v\, v^2 \, \int_{-v^2 T^2}^{0} \diff t\, \tfrac{t}{T^2} 
    \bigg\{\left[3v^2\left(\tfrac{t}{T^2} + v^2\right) +
      4\left(\tfrac{t}{T^2} - 4v^2\right)\right] K_3(v) \\ 
    - 2v\left(\tfrac{3t}{T^2} + 8v^2\right) K_2(v)\bigg\}
    \frac{\diff\sigma}{\diff t}^{ij\rightarrow ij}. 
\end{multline}
We also obtain the inelastic terms as
\begin{equation}
    \Tilde{\omega}[gg] = \frac{x_g^2 N_f}{192} \int_0^{\infty} \diff
    v\, v^6 \left[(v^2+28)K_3(v) - 2v\, K_2(v)\right]
    \sigma^{gg\rightarrow q\bar{q}}(v),
\label{wgg}
\end{equation}
\begin{multline}
\Tilde{\omega}[gq] = \frac{x_g^2}{768} \int_0^{\infty} \diff v\, v^6
\left[(3v^2+4)K_3(v) - 6v\, K_2(v)\right] \sigma_{\rm tr}^{gg\rightarrow
  q\bar{q}}(v) \\ 
- \frac{x_g^2}{384} \int_0^{\infty} \diff v\, v^6
\left [ (v^2+28)K_3(v) - 2v\, K_2(v) \right ] 
\sigma^{gg\rightarrow  q\bar{q}}(v), 
\end{multline}
\begin{equation}
    \Tilde{\omega}[qq] = \frac{x_q
      x_{\bar{q}}}{384} \int_0^{\infty} \diff v\, v^6
    \left[(v^2+48)K_3(v) + 8v\, K_2(v)\right] 
    \left[\sigma^{q\bar{q}\rightarrow gg}(v) +
      (N_f-1)\sigma^{q\bar{q}\rightarrow q^{\prime} \bar{q}^{\prime}}(v)\right], 
\end{equation}
\begin{equation}
    \Tilde{\omega}[q\bar{q}] = \frac{x_q x_{\bar{q}}}{384}
    \int_0^{\infty} \diff v\, v^6 \left[(v^2+8)K_3(v) - 12 v\,
      K_2(v)\right] \left[\sigma^{q\bar{q}\rightarrow gg}(v) +
      (N_f-1)\sigma^{q\bar{q}\rightarrow q^{\prime} \bar{q}^{\prime}}(v)\right], 
\end{equation}
\begin{multline}
    \Tilde{\omega}[qq^{\prime}] = \frac{x_q x_{\bar{q}}}{768}
    \int_0^{\infty} \diff v\, v^6 \left[(3v^2+4)K_3(v) - 6v\,
      K_2(v)\right] \sigma_{\rm tr}^{q\bar{q}\rightarrow q^{\prime}
      \bar{q}^{\prime}}(v) \\  
    - \frac{x_q x_{\bar{q}}}{384} \int_0^{\infty} \diff v\, v^6
    \left[(v^2+28)K_3(v) - 2v\, K_2(v)\right]
    \sigma^{q\bar{q} \rightarrow q^{\prime} \bar{q}^{\prime}}(v). 
\label{wqqprime}
\end{multline}
Note that we have made use of the standard symmetries in pQCD, e.g.,
the facts that $d\sigma^{gq_i\rightarrow gq_i}/dt=d\sigma^{g 
  {\bar{q}_j}\rightarrow g{\bar q}_j}/dt \equiv d\sigma^{gq\rightarrow
  gq}/dt$ and $d\sigma^{gg\rightarrow q_i {\bar     q}_i}/dt \equiv
d\sigma^{gg \rightarrow q {\bar q}}/dt$ are independent of quark
flavor for massless partons. Also note that the total cross section
and the total transport cross section in the above relations are given
by   
\begin{equation}
\sigma^{ij \rightarrow kl}(v)=
\frac{1}{1+\delta_{kl}} \int_{-s}^0 {\frac {d\sigma}{dt}}^{ij \rightarrow
    kl} dt, ~~ \sigma_{\rm tr}^{ij \rightarrow
  kl}(v)=\frac{1}{1+\delta_{kl}} \int_{-s}^0 {\frac {d\sigma}{dt}}^{ij
  \rightarrow kl} \sin^2\!\theta_{\rm cm} \, dt  
\label{dsigmadt}
\end{equation}
respectively, where $\theta_{\rm cm}$ is the scattering angle in the
two-parton center of mass frame. 
In chemical equilibrium, we find $\Tilde{\omega}[gq]
=\Tilde{\omega}[qg] = \Tilde{\omega}[g\bar{q}]
=\Tilde{\omega}[\bar{q}g]$, 
$\Tilde{\omega}[qq] = \Tilde{\omega}[\bar{q}\bar{q}]$,
$\Tilde{\omega}[q\bar{q}]= \Tilde{\omega}[\bar{q} q]$,  
and $C_{ij}^{00}=C_{ji}^{00}$. 

After inverting the matrix ${\mathbf C}$, we use Eq.\eqref{etaVector} to
obtain the general expression of the shear viscosity of a
multi-species massless quark-gluon gas in chemical equilibrium as
\begin{equation}
\eta = 160\, T ~ \frac{x_g^2 \left[C_{qq}^{00} + C_{q\bar{q}}^{00} +
    2(N_f-1)C_{qq^{\prime}}^{00}\right] - 4 N_f x_g x_q\, C_{gq}^{00} +2 N_f
  x_q^2 \,C_{gg}^{00}} {C_{gg}^{00}\left[C_{qq}^{00} +
    C_{q\bar{q}}^{00} + 2(N_f-1)C_{qq^{\prime}}^{00}\right] - 2N_f
  \left(C_{gq}^{00}\right)^2}.
\label{etaNf}
\end{equation}
Note that in chemical equilibrium $x_g=8/(8+6N_f)$ 
and $x_q=x_{\bar q}=3 (1-\delta_{0N_f})/(8+6N_f)$.   
Also note that for the $N_f=0$ case, Eq.\eqref{etaVector} as well as
Eq.\eqref{etaNf} leads to
$\eta=160T/C_{gg}^{00}=160T/\Tilde{c}_{0}[g]$, which as expected is
the Chapman-Enskog result for a single species of massless 
parton~\cite{Plumari:2012ep,MacKay:2022uxo}.  
Interestingly, the explicit expression (i.e., after the inversion of
matrix ${\mathbf C}$) of $\eta$ in Eq.~\eqref{etaNf} involves the
matrix elements $C_{ij}^{00}$ at most to second order, not to the
order of the dimension of the matrix $(1+2N_f)$; this is a result
of the symmetry properties of matrix ${\mathbf C}$ and vector 
$\nicevec{\gamma}$. 

\section{The single-species limits}
\label{limit}

In certain limits, which we call the single-species limits, the
quark-gluon gas should behave in the same way as a single species in
terms of momentum transfer and thus shear viscosity. 
So we can use the previously known Chapman-Enskog result for a
single particle species to check the shear viscosity result for multiple
parton species such as Eq.\eqref{etaNf}. 
One such limit occurs when all the parton species scatter elastically
(but not inelastically) with the same cross section and angular
distribution, i.e., when $d\sigma^{ij\rightarrow
  ij}/dt=(1+\delta_{ij}) d\sigma/dt$, where $d\sigma/dt$ is a
given arbitrary angular distribution. Note that in this case 
\begin{equation}
\sigma_{\rm tr}^{ii\rightarrow
  ii}(v)=\sigma_{\rm tr}^{ij\rightarrow ij}(v)=\int_{-s}^0 
\sin^2\!\theta_{\rm cm} \frac{d\sigma}{dt} dt \equiv \sigma_{\rm tr}(v)
\end{equation}
are independent of parton species ($i$ and $j$), so we call this the
equal-cross section case of the single-species limit. We now check
Eq.\eqref{etaNf} in this limit, where we find the following matrix
elements:
\begin{eqnarray}
&C_{gg}^{00}=x_g^2 \, \Tilde{c}_{0}[g]+x_g(1-x_g) \, \Tilde{c}_{1}[gq],
~C_{qq}^{00}=x_q^2 \,\Tilde{c}_{0}[q]+x_q(1-x_q) \,\Tilde{c}_{1}[gq],
                   \nonumber\\ 
&C_{gq}^{00}=x_g x_q \,\Tilde{c}_{2}[gq],
~C_{q\bar{q}}^{00}=x_q^2 \,\Tilde{c}_{2}[gq],
~C_{qq^{\prime}}^{00}=x_q^2 \,\Tilde{c}_{2}[gq].
\end{eqnarray}
In the above, we have taken advantage of the facts that 
all the inelastic $\Tilde{\omega}[ij]$ terms are zero 
and that $\Tilde{c}_{0}[i], \Tilde{c}_{1}[ij]$ and $\Tilde{c}_{2}[ij]$
are all independent of parton species. 
We have also used the relations $x_q=x_{\bar q}$ and
$\sum x_i=1$ that were imposed in deriving Eq.\eqref{etaNf}.  
Noting $\Tilde{c}_{0}[i]=\Tilde{c}_{1}[ij]+\Tilde{c}_{2}[ij]$,
Eq.\eqref{etaNf} then gives 
\begin{equation}
\eta =\frac{61440\, T}{\int_0^{\infty} \diff v\, v^6 \left[(3v^2 +
    4)K_3(v) - 6v\, K_2(v)\right] \sigma_{\rm tr}(v)},
\label{singleSpecies}
\end{equation}
which is exactly the single-species
result~\cite{Plumari:2012ep,MacKay:2022uxo}.

For $\eta$ expressions that are more explicit than
Eq.\eqref{etaNf} and written in terms of cross sections, 
let us consider the special case of
isotropic and energy($s$)-independent (or constant) scatterings, where
$d\sigma^{ij\rightarrow kl}/dt$ is a constant that can depend on 
$i,j,k,l$ but not the Mandelstam variables. For this case, we get the
following shear viscosity expression for a massless quark-gluon gas of
$N_f$ quark flavors in chemical equilibrium: 
\begin{multline}
\eta^{\rm iso}=\frac {12T}{5} \frac{A}{B}, 
~A=240 (2N_f \sigma_{gg} + \sigma_4) +8(112 + 48 N_f + 63 N_f^2) 
\sigma_{gq} +9 (48 + 32N_f + 27 N_f^2) \sigma_{q\bar q}^{gg}, \\ 
\mspace{-25mu} B = 8 \left [8 \sigma_{gq}(28 \sigma_{gg} + 27N_f
\sigma_{gq})  + 3 \sigma_4 (20 \sigma_{gg} +  21 N_f \sigma_{gq}) \right ]  \\
+ 9 (96 \sigma_{gg} + 176 N_f \sigma_{gq}+27 N_f
\sigma_4) \sigma_{q\bar q}^{gg} +   351 N_f (\sigma_{q\bar q}^{gg})^2. 
\label{etaIso}
\end{multline}
For brevity, we have written the elastic cross section
$\sigma^{ij\rightarrow   ij}$ as $\sigma_{ij}$, the two independent
inelastic cross sections as $\sigma_{q\bar q}^{gg} \equiv \sigma(q\bar
q \rightarrow gg)$ and $\sigma_{q\bar q}^{q^{\prime} \bar q^{\prime}}
\equiv \sigma(q\bar q \rightarrow q^{\prime} \bar   q^{\prime})$,
respectively, and 
\begin{equation}
\sigma_4 \equiv \sigma_{qq} + \sigma_{q\bar q}+ (N_f-1)
(2\sigma_{qq^{\prime}} + \sigma_{q\bar q}^{q^{\prime} \bar q^{\prime}}).  
\end{equation}
Note that $\sigma_{gg}^{q\bar q} \equiv \sigma(gg \rightarrow q\bar
q)=9 \,\sigma(q\bar q \rightarrow gg)/32$; thus it is not
independent.

For this case of isotropic and constant cross sections, 
in the limit that the collision rate of every parton species per
parton is the same, the shear viscosity should be given by the
single-species result $\eta_1^{\rm iso}=6T/(5\sigma)$. 
For the collision rate $r_i$ of parton species $i$ per parton, we have 
\begin{eqnarray}
r_g/n=x_g \sigma_{gg} + 2 N_f x_q \sigma_{gq} + N_f x_g
\sigma_{gg}^{q\bar q}, ~r_q/n=r_{\bar q}/n=x_g \sigma_{gq} + 
x_q \sigma_4 +  x_q \sigma_{q\bar q}^{gg}.
\end{eqnarray}
Thus the equal rate condition $r_g=r_q$ requires the following
relation among the seven independent cross sections:
\begin{equation}
\sigma_{gg} = \frac {3 }{8}\sigma_4 +\frac {(4-3 N_f)}{32} \left ( 8
  \sigma_{gq}  + 3 \sigma_{q\bar q}^{gg} \right ),
\label{sigmagg}
\end{equation}
then Eq.\eqref{etaIso} reduces to 
\begin{equation} 
\eta=\frac{6 (4 + 3 N_f) T} {5 (4 \sigma_{gg} + 3 N_f \sigma_{gq} +
  4 N_f \sigma_{gg}^{q\bar q})}.
\label{etaNfiso}
\end{equation}
On the other hand, the quark-gluon gas in this equal-rate limit
behaves, in terms of momentum transfer, as a single parton species
with an effective isotropic cross section $\sigma_{\rm eff}$ that gives the
same collision rate per parton. From  $n\sigma_{\rm eff}=r_g$, we get the
equivalent effective cross section as 
\begin{equation} 
\sigma_{\rm eff}=\frac{4\sigma_{gg} + 3N_f \sigma_{gq} + 4 N_f
  \sigma_{gg}^{q\bar  q} }{4 + 3 N_f}.
\label{sigmaIso}
\end{equation}
The single-species result $\eta_1^{\rm iso}=6T/(5\sigma)$ 
for $\sigma=\sigma_{\rm eff}$ thus agrees with Eq.\eqref{etaNfiso},
i.e., the shear viscosity for isotropic and constant cross
sections in Eq.\eqref{etaIso} satisfies this single-species limit. 

We can extend the above cases to a general single-species limit, 
where $d\sigma^{ij\rightarrow kl}/dt=f_{ij}^{kl} (1+\delta_{kl})
d\sigma_0/dt$ with $d\sigma_0/dt$ being a given arbitrary angular
distribution and $f_{ij}^{kl}$ being constants that depend on 
$i,j,k,l$ but not on the Mandelstam variables. 
When the seven independent $\sigma^{ij\rightarrow kl}(v)$ values
satisfy Eq.\eqref{sigmagg}, then each parton species has the same 
collision rate per parton and the general Eq.\eqref{etaNf} indeed
reduces to the single species result of Eq.\eqref{singleSpecies}.  
We note that the previous two single-species limits in this Section are  
simply special cases of this general single-species limit.  
The proof of this general equal-rate case of the single-species limit
is shown in Appendix C, which serves as a strong check of the 
general expression of shear viscosity in Eq.\eqref{etaNf}. 

\section{Discussions}
\label{discussion}

In a follow-up study~\cite{Ohanaka:2026gla}, we applied the general
shear viscosity expression of Eq.\eqref{etaNf} to given parton cross
sections at finite temperature based on perturbative QCD. We find that
the Chapman-Enskog results of shear viscosity are 
qualitatively similar to the corresponding leading-order results from
the AMY  framework~\cite{Arnold:2000dr,Arnold:2003zc,Moore:2025}; 
quantitatively they are higher partly due to the usage of parton
thermal masses (instead of self-energies) to screen the cross
sections. It will also be very useful to compare the Chapman-Enskog
results for given parton cross sections with numerical results of
shear viscosity obtained from transport models using the Green-Kubo
relation. In addition, it will be interesting to consider the case of finite 
parton mass including different masses of different particle species; 
we plan to carry out this separate study in the near future. 

This study includes inelastic two-body cross sections in the
Chapman-Enskog calculation of shear viscosity for the first time. The
effects of these inelastic processes have been shown in
follow-up studies~\cite{Ohanaka:2026gla,Ross:2026qxu} 
for several choices of the renormalization scale $Q$ used in the
strong coupling constant and the scaling factor $\xi$ used for
screening the scattering matrix elements.  For $N_f=3$, inelastic
two-body scatterings reduce the shear viscosity by $\sim 5-12\%$
depending on the temperature as well as the values of $Q$ and
$\xi$. We also note that this study only considers Boltzmann
statistics but not quantum statistics. We have compared the $\eta/s$ 
values~\cite{Ohanaka:2026gla} between Boltzmann statistics and quantum  
statistics from the leading-order AMY framework~\cite{Moore:2025},
where the shear viscosity at $N_f=3$ for two-body scatterings from
Boltzmann statistics is higher than that from quantum statistics by
$\sim 13-15\%$. Therefore, the effect from quantum statistics needs to
be considered in the future for quantitative comparisons between
parton transport results and the experimental data. 

\section{Conclusion}
\label{conclusion}

We have calculated the shear viscosity of a massless quark-gluon
gas in chemical equilibrium subjected to any
$2\leftrightarrow 2$ interactions, including all elastic and inelastic
collisions with arbitrary angular distributions. 
The general result is an explicit first-order Chapman-Enskog
expression for the shear viscosity for a system of gluons and  $N_f$
flavors of quarks with the Boltzmann statistics. As expected, the
general result for a quark-gluon gas reduces to the previously known
result for a massless single particle species in the equal-cross
section case, where all the parton species scatter elastically with
the same cross section and angular distribution without inelastic
scatterings.  We then design a general single-species limit, 
where the shear viscosity of the quark-gluon system is expected to
reduce to the single species result. 
In this general limit, all the parton species scatter with the same
angular distribution but can have different cross section magnitudes
and scatter both elastically and inelastically, and the seven
independent cross sections satisfy one relation so that the collision
rate of each parton species is the same. 
In this limit, which can be called the equal-rate case of the
single-species limit, we show that the general expression of the shear
viscosity indeed reduces to the single-species result.  
This provides a strong check of the general expression  
because the general single-species limit checks not only the elastic
terms but also the inelastic terms in the collision matrix. 
In addition, we give an explicit expression of the shear viscosity in
terms of all the $2\leftrightarrow 2$ cross sections for the case of
isotropic and energy-independent scatterings, where it is much easier
to see how different processes contribute to the shear viscosity. 
These analytical results of the shear viscosity should be useful for 
parton transport models and its extraction of shear viscosity from 
comparisons with the experimental data. 
They can also be applied to QCD cross sections to facilitate studies of 
the shear viscosity of quark-gluon matter at finite temperature
with parton transport models or QCD effective kinetic theory.

\section*{Acknowledgments}
We thank Drs. P.B. Arnold, V. Greco, G.D. Moore, S. Plumari and
X.N. Wang for helpful discussions. This work has been supported by the
National Science Foundation under Grant No. 2310021.

\appendix
\section{Derivation of Eq.(15)}

We start from the following Eq.(8):
\begin{equation}
\sum_{j=1}^N x_j R_{ij}[C_i(\tau_i)\pi\traceless{{}_i^\mu \pi_i^\nu}]
=\pi\traceless{{}_i^\mu \pi_i^\nu}. 
\end{equation}
Let us apply the inner product $\langle F, x_i \tau_i^r
\pi\traceless{_{i\mu} \pi_{i\nu}} \rangle_i$ to the above equation, 
where $F$ represents either side of the equation. 
When $F$ is replaced with the right-hand-side, the above procedure 
simply gives $x_i \gamma_{i,r}$, where $\gamma_{i,r} \equiv
\big\langle\tau_i^r \pi\traceless{_i^{\mu} \pi_i^{\nu}} ,
\pi\traceless{_{i\mu} \pi_{i\nu}} \big\rangle_i$. 
When $F$ is replaced with the left-hand-side, the above procedure
gives four terms since $R_{ij}[\phi_i] \propto (\phi_i + \phi_j -
\phi_k - \phi_l)$ as shown in Eq.(9). 
The first term ($T_1$) is given by
\begin{equation}
T_1=\frac{1}{128\pi^2 T^6} \sum_{s=0}^P \sum_{j,k,l} \int \frac{\diff^3
  \nicevec{p}_i}{p_i^0} \frac{\diff^3
  \nicevec{p}_j}{p_j^0} \frac{\diff^3 \nicevec{p}_k}{p_k^0}
\frac{\diff^3 \nicevec{p}_l}{p_l^0} e^{-\tau_i -\tau_j}
x_j  C_{i,s}^{(P)} \tau_i^s \pi\traceless{{}_i^\mu \pi_i^\nu}~W_{ij\rightarrow kl} 
~x_i \tau_i^r \pi\traceless{_{i\mu} \pi_{i\nu}}, 
\end{equation}
where we have made the approximation $C_i(\tau_i)
\approx \sum_{s=0}^{P} C_{i,s}^{(P)} \tau_i^s$. 
Writing $C_{i,s}^{(P)}$ as $\sum_m \delta_{im} C_{m,s}^{(P)}$ and then
making the exchange $m \leftrightarrow j$, we get 
\begin{align}
T_1&=\sum_{s=0}^P \sum_j C_{j,s}^{(P)} \frac{\delta_{ij} x_i}{128\pi^2
  T^6} \sum_{k,l,m} x_m \int \frac{\diff^3   \nicevec{p}_i}{p_i^0}
\frac{\diff^3    \nicevec{p}_m}{p_m^0} \frac{\diff^3
  \nicevec{p}_k}{p_k^0} \frac{\diff^3 \nicevec{p}_l}{p_l^0} e^{-\tau_i
  -\tau_m} \tau_i^s \pi\traceless{{}_i^\mu \pi_i^\nu}~W_{im\rightarrow
  kl}  ~\tau_i^r \pi\traceless{_{i\mu} \pi_{i\nu}} \\
&=\sum_{s=0}^P \sum_j C_{j,s}^{(P)} \delta_{ij} x_i \sum_{k,l,m} x_m [\tau_i^r
\pi\traceless{_i^{\mu} \pi_i^{\nu}} \, , \; \tau_i^s
\pi\traceless{_{i\mu} \pi_{i\nu}} ]_{im\rightarrow kl}.
\end{align}
We see that the expression after the term $C_{j,s}^{(P)}$ in the above
is just the first term of the matrix element $ C_{ij}^{rs}$ in
Eq.(13). 
The second term of the left-hand-side from the above procedure is
simpler and given by
\begin{equation}
T_2=\sum_{s=0}^P \sum_j C_{j,s}^{(P)} x_i x_j \sum_{k,l} [\tau_i^r
\pi\traceless{_i^{\mu} \pi_i^{\nu}} \, , \; \tau_j^s
\pi\traceless{_{j\mu} \pi_{j\nu}} ]_{ij\rightarrow kl}, 
\end{equation}
which contains the second term of the matrix element $ C_{ij}^{rs}$.

The third term is given by
\begin{equation}
T_3=\frac{-1}{128\pi^2 T^6} \sum_{s=0}^P \sum_{j,k,l} \int \frac{\diff^3
  \nicevec{p}_i}{p_i^0} \frac{\diff^3
  \nicevec{p}_j}{p_j^0} \frac{\diff^3 \nicevec{p}_k}{p_k^0}
\frac{\diff^3 \nicevec{p}_l}{p_l^0} e^{-\tau_i -\tau_j}
x_j  C_{k,s}^{(P)} \tau_k^s \pi\traceless{{}_k^\mu \pi_k^\nu}~W_{ij\rightarrow kl} 
~x_i \tau_i^r \pi\traceless{_{i\mu} \pi_{i\nu}}.
\end{equation}
Making the exchange $j \leftrightarrow k$ and the replacement $l
\rightarrow m$, we get
\begin{equation}
T_3=\frac{-1}{128\pi^2 T^6} \sum_{s=0}^P \sum_{j,k,m} \int \frac{\diff^3
  \nicevec{p}_i}{p_i^0} \frac{\diff^3
  \nicevec{p}_j}{p_j^0} \frac{\diff^3 \nicevec{p}_k}{p_k^0}
\frac{\diff^3 \nicevec{p}_m}{p_m^0} e^{-\tau_i -\tau_k}
x_k  C_{j,s}^{(P)} \tau_j^s \pi\traceless{{}_j^\mu \pi_j^\nu}~W_{ik\rightarrow jm} 
~x_i \tau_i^r \pi\traceless{_{i\mu} \pi_{i\nu}}.
\end{equation}
Because of the detailed balance relation (for massless particles) 
$d_i d_k W_{ik\rightarrow jm}=d_j d_m W_{jm\rightarrow ik}$
with $d_i$ being the degeneracy factor of parton species $i$, we have
$x_i x_k W_{ik\rightarrow jm}=x_j x_m W_{jm\rightarrow ik}$ in chemical
equilibrium. Also noting the energy-momentum conservation relation 
$\tau_i+\tau_k=\tau_j+\tau_m$, we then get 
\begin{align}
T_3&=\frac{-1}{128\pi^2 T^6} \sum_{s=0}^P \sum_{j,k,m} \int \frac{\diff^3
  \nicevec{p}_i}{p_i^0} \frac{\diff^3
  \nicevec{p}_j}{p_j^0} \frac{\diff^3 \nicevec{p}_k}{p_k^0}
\frac{\diff^3 \nicevec{p}_m}{p_m^0} e^{-\tau_j -\tau_m}
x_j  C_{j,s}^{(P)} \tau_j^s \pi\traceless{{}_j^\mu \pi_j^\nu}~W_{jm \rightarrow ik} 
~x_m \tau_i^r \pi\traceless{_{i\mu} \pi_{i\nu}}\\
&=\sum_{s=0}^P \sum_j C_{j,s}^{(P)} x_j \sum_{k,m} x_m [\tau_i^r
\pi\traceless{_i^{\mu} \pi_i^{\nu}} \, , \; \tau_j^s
\pi\traceless{_{j\mu} \pi_{j\nu}} ]_{jm\rightarrow ik},
\end{align}
which contains the third term of the matrix element $ C_{ij}^{rs}$.
Similarly, one can show that the fourth term reduces to the following 
(after taking advantage of $W_{ij\rightarrow kl}=W_{ji\rightarrow lk}$):
\begin{equation}
T_4=\sum_{s=0}^P \sum_j C_{j,s}^{(P)} x_j \sum_{k,m} x_m [\tau_i^r
\pi\traceless{_i^{\mu} \pi_i^{\nu}} \, , \; \tau_j^s
\pi\traceless{_{j\mu} \pi_{j\nu}} ]_{jm\rightarrow ki},
\end{equation}
which contains the forth term of the matrix element $ C_{ij}^{rs}$. 

Adding all four terms from the left-hand-side together, we then obtain
Eq.(15)~\cite{deGroot1980}:
\begin{equation}
\sum_{j=1}^N \sum_{s=0}^P C_{ij}^{rs} C_{j,s}^{(P)} = x_i\gamma_{i,r}.
\end{equation}

\section{Calculating the collision brackets in matrix
${\mathbf C}$}

For completeness, here we show the derivations of the collision brackets
for massless partons, similar to those for massive
partons with elastic collisions in the book~\cite{deGroot1980}. 
For two incoming partons (with momenta $p_i$ and $p_j$) and two outgoing
partons (with momenta $p_k$ and $p_l$), the total 4-momentum is 
\begin{equation}
    P^{\mu} \equiv p_i^{\mu} + p_j^{\mu} = p_k^{\mu} + p_l^{\mu}
    \equiv {P'}^{\mu}, 
\end{equation}
and the projected relative 4-momenta are defined as~\cite{deGroot1980}
\begin{equation}
Q^{\mu} \equiv \Delta_P^{\mu\nu} (p_{i\nu} - p_{j\nu}), \;\;
{Q'}^{\mu} \equiv \Delta_P^{\mu\nu} (p_{k\nu} - p_{l\nu}), 
{\rm ~with~} \Delta_P^{\mu\nu} \equiv g^{\mu\nu} - P^{\mu} P^{\nu}/P^2.
\end{equation}
Taking advantage of the orthogonality relations $P^{\mu} Q_{\mu} =
P^{\mu} Q_{\mu}' = 0$, we can write the individual momentum of each 
massless parton as 
\begin{eqnarray}
p_i^{\mu}=(P^{\mu} + Q^{\mu})/2, ~p_j^{\mu}=(P^{\mu} - Q^{\mu})/2,
~p_k^{\mu} = (P^{\mu} + Q^{\prime \mu})/2, 
~p_l^{\mu} =(P^{\mu} -  Q^{\prime \mu})/2. 
\end{eqnarray}
We can then write a generic collision integral~\cite{deGroot1980},
similar in structure to Eq.\eqref{fgbracket}, which can represent all
the terms in the elements of matrix ${\mathbf C}$: 
\begin{multline}
    J_{ij\rightarrow kl}^{(a,b,d,e,f)} = \frac{1}{128\pi^2 T^6} \int
    \frac{\diff^3 \nicevec{p}_i}{p_i^0} \frac{\diff^3
      \nicevec{p}_j}{p_j^0} \frac{\diff^3 \nicevec{p}_k}{p_k^0}
    \frac{\diff^3 \nicevec{p}_l}{p_l^0} e^{-\frac{P\cdot U}{T}} \\ 
   \times\left(\frac{P^2}{T^2}\right)^a \left(\frac{P\cdot
        U}{T}\right)^b \left(\frac{Q\cdot U}{T}\right)^d
    \left(\frac{Q'\cdot U}{T}\right)^e \left(\frac{-Q\cdot
        Q'}{T^2}\right)^f W_{ij\rightarrow kl},  
\end{multline}
where all the upper indices ($a,b,d,e,f$) are integers. 
Following earlier derivations~\cite{deGroot1980,Moroz:2014gda}, we get
the final result for the $J$ integral for massless partons as
\begin{multline}
    J_{ij\rightarrow kl}^{(a,b,d,e,f)} = \frac{\pi(-1)^b (d+e+1)!!}{8}
\int_0^{\infty} dv\; v^{2a+b+d+e+2f+5} \\
\times \left(\frac{\diff^b}{\diff v^b}
  \frac{K_{(d+e)/2+1}(v)}{v^{(d+e)/2+1}}\right)
\sum_{g=0}^{\mathrm{min}(d,e)} K(d,e,g) \sigma_{ij\rightarrow
  kl}^{(f,g)}(v).
\end{multline}
In the above, $K_n(v)$ is the modified Bessel of the second kind,
\begin{equation}
K(d,e,g) = \frac{d! \, e!}{(d-g)!! \, (d+g+1)!! \, (e-g)!! \, (e+g+1)!!}
\end{equation}
for even $(d-g)$ and even $(e-g)$ while $K(d,e,g)=0$ otherwise, and 
\begin{equation}
    \sigma_{ij\rightarrow kl}^{(f,g)}(v) = \frac{2g+1}{4\pi} \int
    \diff\Omega \, \cos^f\mspace{-5mu}\Theta\; P_g(\cos\Theta)
    \frac{\diff\sigma}{\diff\Omega}^{ij\rightarrow kl},
\end{equation}
where $v=\sqrt{s}/T$ and $P_g(x)$ represents the Legendre
polynomials.

Examining the collision brackets in $C_{ij}^{rs}$ of
Eq.\eqref{eqnCijrs}, we can rewrite the integrands using 
\begin{equation}
    \begin{aligned}
        a\traceless{^{\mu} a^{\nu}} b\traceless{_{\mu} b_{\nu}} &=
        \left(\Delta_{\mu\nu} a^{\mu} b^{\nu}\right)^2 -
        \tfrac{1}{3}\left(\Delta_{\mu\nu} a^{\mu}
          a^{\nu}\right)\left(\Delta_{\alpha\beta} b^{\alpha}
          b^{\beta}\right) \\ 
        &= \left[(a\cdot b)-(a\cdot U)(b\cdot U)\right]^2 -
        \tfrac{1}{3} \left[a^2 - (a\cdot U)^2\right] \left[b^2 -
          (b\cdot U)^2\right]. 
    \end{aligned}
\end{equation} 
For massless partons, the first term on the right-hand-side of
Eq.\eqref{eqnCijrs} corresponds to the following integrals:
\begin{equation}
    [\tau_i^r \pi\traceless{_i^{\mu} \pi_i^{\nu}} \, , \; \tau_i^s
    \pi\traceless{_{i\mu} \pi_{i\nu}} ]_{im\rightarrow kl} =
    \frac{2}{3} [\tau_i^{r+2},\tau_i^{s+2}]_{im\rightarrow kl}, 
\end{equation}
\begin{equation}
    [\tau_i^r\, , \; \tau_i^s ]_{im\rightarrow kl} = \frac{1}{2^{r+s}}
    \sum_{n=0}^{r+s} \binom{r+s}{n} J_{im\rightarrow
      kl}^{(0,n,r+s-n,0,0)}. 
\end{equation}
The second term in Eq.\eqref{eqnCijrs} corresponds to the
$s-$channel integrals: 
\begin{multline}
    [\tau_i^r \pi\traceless{_i^{\mu} \pi_i^{\nu}} \, , \; \tau_j^s
    \pi\traceless{_{j\mu} \pi_{j\nu}} ]_{ij\rightarrow kl} =  
    -2 [\tau_i^{r+1} \pi\overline{_i^{\mu}},\tau_j^{s+1}
    \pi\overline{_{j\mu}}]_{ij\rightarrow kl} 
- \frac{4}{3}[\tau_i^{r+2},\tau_j^{s+2}]_{ij\rightarrow kl} \\ 
    + \frac{1}{2^{r+s+2}} \sum_{n_1=0}^r \sum_{n_2=0}^s (-1)^{s-n_2}
    \binom{r}{n_1} \binom{s}{n_2} J_{ij\rightarrow
      kl}^{(2,n_1+n_2,r+s-n_1-n_2,0,0)},  
\end{multline}
\begin{multline}
    [\tau_i^r \pi\overline{_i^{\mu}} \, , \; \tau_j^s
    \pi\overline{_{j\mu}} ]_{ij\rightarrow kl} =  
    - [\tau_i^{r+1},\tau_j^{s+1}]_{ij\rightarrow kl} \\
    + \frac{1}{2^{r+s+1}} \sum_{n_1=0}^r \sum_{n_2=0}^s (-1)^{s-n_2}
    \binom{r}{n_1} \binom{s}{n_2} J_{ij\rightarrow
      kl}^{(1,n_1+n_2,r+s-n_1-n_2,0,0)},
\end{multline}
\begin{equation}
    [\tau_i^r\, , \; \tau_j^s ]_{ij\rightarrow kl} = \frac{1}{2^{r+s}}
    \sum_{n_1=0}^r \sum_{n_2=0}^s (-1)^{s-n_2} \binom{r}{n_1}
    \binom{s}{n_2} J_{ij\rightarrow
      kl}^{(0,n_1+n_2,r+s-n_1-n_2,0,0)}. 
\end{equation}
In the above, $\pi\overline{_i^{\mu}} \equiv
\Delta^{\mu}_{\nu} \pi_i^{\nu}$,  where $\Delta^{\mu\nu}$ has been
defined in Eq.\eqref{deltamunu}. 
 
The third term in Eq.\eqref{eqnCijrs} corresponds to the
$t-$channel integrals:
\begin{multline}
    [\tau_i^r \pi\traceless{_i^{\mu} \pi_i^{\nu}}\, , \;  \tau_j^s
    \pi\traceless{_{j\mu} \pi_{j\nu}}]_{jm\rightarrow ik} =  
    -2 [\tau_i^{r+1} \pi\overline{_i^{\mu}},\tau_j^{s+1}
    \pi\overline{_{j\mu}}]_{jm\rightarrow ik} 
-   \frac{4}{3}[\tau_i^{r+2},\tau_j^{s+2}]_{jm\rightarrow ik} \\ 
    +\frac{1}{2^{r+s+4}} \sum_{n_1=0}^r \sum_{n_2=0}^s \binom{r}{n_1}
    \binom{s}{n_2}     \left [ J_{jm\rightarrow
      ik}^{(2,n_1+n_2,s-n_2,r-n_1,0)} \right . \\ 
\left . - 2  J_{jm\rightarrow ik}^{(1,n_1+n_2,s-n_2,r-n_1,1)} +
    J_{jm\rightarrow ik}^{(0,n_1+n_2,s-n_2,r-n_1,2)} \right ],
\end{multline}
\begin{multline}
    [\tau_i^r \pi\overline{_i^{\mu}} \, , \; \tau_j^s
    \pi\overline{_{j\mu}} ]_{jm\rightarrow ik} = -
    [\tau_i^{r+1},\tau_j^{s+1}]_{jm\rightarrow ik} \\
    + \frac{1}{2^{r+s+2}} \sum_{n_1=0}^r \sum_{n_2=0}^s \binom{r}{n_1}
    \binom{s}{n_2} \left [  J_{   jm\rightarrow 
      ik}^{(1,n_1+n_2,s-n_2,r-n_1,0)}  
-   J_{jm\rightarrow ik}^{(0,n_1+n_2,s-n_2,r-n_1,1)}\right ], 
\end{multline}
\begin{equation}
    [\tau_i^r,\tau_j^s]_{jm\rightarrow ik} = \frac{1}{2^{r+s}}
    \sum_{n_1=0}^r \sum_{n_2=0}^s  \binom{r}{n_1} \binom{s}{n_2}
    J_{jm\rightarrow ik}^{(0,n_1+n_2,s-n_2,r-n_1,0)}. 
\end{equation}

The fourth term in Eq.\eqref{eqnCijrs} corresponds to the 
$u-$channel integrals: 
\begin{multline}
    [\tau_i^r \pi\traceless{_i^{\mu} \pi_i^{\nu}}\, , \;  \tau_j^s
    \pi\traceless{_{j\mu} \pi_{j\nu}}]_{jm\rightarrow ki} =   
    -2 [\tau_i^{r+1} \pi\overline{_i^{\mu}},\tau_j^{s+1}
    \pi\overline{_{j\mu}}]_{jm\rightarrow ki} 
-    \frac{4}{3}[\tau_i^{r+2},\tau_j^{s+2}]_{jm\rightarrow ki} \\ 
    + \frac{1}{2^{r+s+4}} \sum_{n_1=0}^r \sum_{n_2=0}^s (-1)^{r-n_1}
   \binom{r}{n_1}  \binom{s}{n_2} \left [ J_{jm\rightarrow
      ki}^{(2,n_1+n_2,s-n_2,r-n_1,0)} \right . \\ 
\left . + 2  J_{jm\rightarrow ki}^{(1,n_1+n_2,s-n_2,r-n_1,1)} +
    J_{jm\rightarrow  ki}^{(0,n_1+n_2,s-n_2,r-n_1,2)}\right ], 
\end{multline}
\begin{multline}
    [\tau_i^r \pi\overline{_i^{\mu}} \, , \; \tau_j^s
    \pi\overline{_{j\mu}} ]_{jm\rightarrow ki} = -
    [\tau_i^{r+1},\tau_j^{s+1}]_{jm\rightarrow ki} \\ 
    + \frac{1}{2^{r+s+2}} \sum_{n_1=0}^r  \sum_{n_2=0}^s (-1)^{r-n_1}
    \binom{r}{n_1} \binom{s}{n_2} \left [ J_{jm\rightarrow
      ki}^{(1,n_1+n_2,s-n_2,r-n_1,0)}+
    J_{jm\rightarrow ki}^{(0,n_1+n_2,s-n_2,r-n_1,1)}\right ], 
\end{multline}
\begin{equation}
    [\tau_i^r,\tau_j^s]_{jm\rightarrow ki} = \frac{1}{2^{r+s}}
    \sum_{n_1=0}^r \sum_{n_2=0}^s (-1)^{r-n_1} \binom{r}{n_1}
    \binom{s}{n_2} J_{jm\rightarrow ki}^{(0,n_1+n_2,s-n_2,r-n_1,0)}. 
\end{equation}

\section{Verification of the first-order result in a general
  single-species limit} 

Here we prove that the shear viscosity expression in Eq.\eqref{etaNf}
satisfies the general single-species limit, where one can write 
\begin{equation}
\frac {d\sigma^{ij\rightarrow kl}}{dt}=f_{ij}^{kl} (1+\delta_{kl})
\frac {d\sigma_0}{dt} 
\end{equation}
and each parton species has the same collision rate per parton. 
In the above, $d\sigma_0/dt$ is an arbitrary given differential cross
section, and $f_{ij}^{kl}$ are constants that depend on $i,j,k,l$
but not on the Mandelstam variables.   
As a result of Eq.\eqref{dsigmadt}, we then have
\begin{equation}
\sigma_{ij}^{kl} \equiv \sigma^{ij\rightarrow kl}(v)=f_{ij}^{kl}
\int_{-s}^0 \frac {d\sigma_0}{dt} dt \equiv f_{ij}^{kl} \, \sigma_0(v), 
\end{equation}
where $v=\sqrt{s}/T$, thus the cross section of every $2 \rightarrow
2$ scattering channel has the same energy($s$)-dependence and is
proportional to the $f_{ij}^{kl}$ constant. For elastic scatterings,
we shall write  $\sigma^{ij\rightarrow   ij}(v)$ as $\sigma_{ij}$ and
$f_{ij}^{ij}$ as $f_{ij}$ in the following. 

We can write the following matrix elements that Eq.\eqref{etaNf} needs:
\begin{eqnarray}
&C_{gg}^{00}=x_g^2 \, \Tilde{c}_{0}[g]+x_g(1-x_g) \,
                   \Tilde{c}_{1}[gq]+ \Tilde{\omega}[gg], \nonumber\\ 
&C_{qq}^{00}=x_q^2 \,\Tilde{c}_{0}[q]+x_q \big [ x_g \,\Tilde{c}_{1}[gq] +x_q \,\Tilde{c}_{1}[q\bar q] +2(N_f-1)x_q \,\Tilde{c}_{1}[qq^{\prime}] \, \big ] + \Tilde{\omega}[qq],                    \nonumber\\ 
&C_{gq}^{00}=x_g x_q \,\Tilde{c}_{2}[gq]+ \Tilde{\omega}[gq],
~C_{q\bar{q}}^{00}=x_q^2 \,\Tilde{c}_{2}[q\bar{q}]+ \Tilde{\omega}[q\bar{q}],
~C_{qq^{\prime}}^{00}=x_q^2 \,\Tilde{c}_{2}[qq^{\prime}]+ \Tilde{\omega}[qq^{\prime}],
\end{eqnarray}
where according to Eq.\eqref{c0g} we have
\begin{equation}
\Tilde{c}_{0}[g]= \frac{f_{gg}}{384} \int_0^{\infty}
\diff v\, v^6 \left [ (3v^2+4)K_3(v) - 6v\,
      K_2(v) \right ] \sigma_{0,\rm tr}(v) \equiv f_{gg}\, \Tilde{c}_{0}.
\label{c0}
\end{equation}
In the above, $\sigma_{0,\rm tr}(v)=\int_{-s}^0 dt \sin^2\!\theta_{\rm
  cm} \, d\sigma_0/dt$. We then write the inelastic terms as
\begin{eqnarray}
&\Tilde{\omega}[gg]=\sigma_{gg}^{q\bar q} \, w_1,
~\Tilde{\omega}[gq]=\sigma_{gg}^{q\bar q} \, w_2,
~\Tilde{\omega}[qq]=\left [\sigma_{q\bar q}^{gg} +(N_f-1) \sigma_{q\bar
  q}^{q^{\prime} \bar q^{\prime}} \right ] w_3, \nonumber \\
&\Tilde{\omega}[q\bar q]=\left [\sigma_{q\bar q}^{gg} +(N_f-1) \sigma_{q\bar q}^{q^{\prime} \bar q^{\prime}} \right ] w_4,
~\Tilde{\omega}[q q^{\prime}]=\sigma_{q\bar q}^{q^{\prime} \bar q^{\prime}} w_5,
\label{omega}
\end{eqnarray}
where the dimensionless $w_1$ to $w_5$ are defined according to
Eqs.\eqref{wgg}-\eqref{wqqprime} as the following: 
\begin{equation}
w_1= \frac{x_g^2 N_f}{192 \, \sigma_0(v)} \int_0^{\infty}
\diff v\, v^6 \left[(v^2+28)K_3(v) - 2v\, K_2(v)\right] \sigma_0(v), 
\end{equation}
\begin{multline}
w_2= \frac{x_g^2}{768 \, \sigma_0(v)} \int_0^{\infty} \diff v\, v^6
\left[(3v^2+4)K_3(v) - 6v\, K_2(v)\right] \sigma_{0,\rm tr}(v)\\
- \frac{x_g^2}{384 \, \sigma_0(v)} \int_0^{\infty} \diff v\, v^6
\left [ (v^2+28)K_3(v) - 2v\, K_2(v) \right ] \sigma_0(v), 
\end{multline}
\begin{equation}
w_3= \frac{x_q^2}{384 \, \sigma_0(v)} \int_0^{\infty} \diff v\, v^6
    \left[(v^2+48)K_3(v) + 8v\, K_2(v)\right] \sigma_0(v),
\end{equation}
\begin{equation}
w_4= \frac{x_q^2}{384 \, \sigma_0(v)}
    \int_0^{\infty} \diff v\, v^6 \left[(v^2+8)K_3(v) - 12 v\,
      K_2(v)\right] \sigma_0(v),
\end{equation}
\begin{multline}
 w_5= \frac{x_q^2}{768 \, \sigma_0(v)}
    \int_0^{\infty} \diff v\, v^6 \left[(3v^2+4)K_3(v) - 6v\,
      K_2(v)\right] \sigma_{0,\rm tr}(v) \\  
    - \frac{x_q^2}{384 \, \sigma_0(v)} \int_0^{\infty} \diff v\, v^6
    \left[(v^2+28)K_3(v) - 2v\, K_2(v)\right]  \sigma_0(v). 
\end{multline}
Note that although $w_k$ ($k=1$ to 5) above has a $v$-dependence due 
to the $\sigma_0(v)$ term in its denominator, $\Tilde{\omega}[ij]$ in 
Eq.\eqref{omega} is independent of $v$ because the $v$-dependence of
$\sigma_0(v)$ is canceled by that of $\sigma^{ij\rightarrow kl}(v)$.

For the limit under consideration here, we find the following
relations:
\begin{eqnarray}
\Tilde{c}_{0}[g]=\frac{\sigma_{gg}}{\sigma_{gq}}
  \left (\Tilde{c}_1[gq]+\Tilde{c}_2[gq] \right ), 
~\frac{\Tilde{c}_{0}[q]}{\Tilde{c}_{0}[g]}=\frac{\sigma_{qq}}{\sigma_{gg}}, 
~\frac{\Tilde{c}_{i}[q\bar q]}{\Tilde{c}_{i}[gq]}=\frac{ \sigma_{q\bar
  q}}{\sigma_{gq}}, 
~\frac{\Tilde{c}_{i}[q q^{\prime}]}{\Tilde{c}_{i}[gq]}=\frac{\sigma_{q
  q^{\prime}}}{\sigma_{gq}},
\end{eqnarray}
where $i=1$ or 2. Then we can write the shear viscosity of
Eq.\eqref{etaNf} in terms of $\Tilde{c}_1[gq],\Tilde{c}_2[gq]$ and
$w_k$, where $w_3$ and $w_4$ always appear as $(w_3+w_4)$.  
We also find the following relations:
\begin{equation}
w_1 =\frac{N_f x_g^2}{x_q^2} \left (w_3+w_4 \right ), 
~w_2 =\frac{x_g^2}{x_q^2} \, w_5, 
~\frac{\Tilde{c}_{0}[g]}{\sigma_{gg}}=\frac{w_1}{N_f
  x_g^2}+\frac{2w_2}{x_g^2}. 
\end{equation}
which we then use to write the shear viscosity of Eq.\eqref{etaNf} in
terms of $\Tilde{c}_1[gq],\Tilde{c}_2[gq]$ and $w_1$. 

For the collision rates per parton, we have 
\begin{eqnarray}
&r_g/n=x_g {\bar \sigma}_{gg} + 2 N_f x_q {\bar \sigma}_{gq} + N_f x_g
  {\bar \sigma}_{gg}^{q\bar q}, \nonumber \\
& r_q/n \!=\! r_{\bar q}/n \!=\! x_g {\bar \sigma}_{gq} + x_q {\bar \sigma}_{qq} 
+ x_q {\bar \sigma}_{q\bar q} + 2 (N_f - 1) x_q {\bar \sigma}_{q q^{\prime}}
+ (N_f - 1) x_q {\bar \sigma}_{q\bar q}^{q^{\prime} \bar q^{\prime}}
+  x_q {\bar \sigma}_{q\bar q}^{gg}.   
\end{eqnarray}
Here we have use the notation $\bar {\sigma} \equiv \langle \sigma
v_{\rm   rel}\rangle$ for the thermal average of a cross section, with
the relative velocity between two colliding massless partons given
by  $v_{\rm rel}=s/(2E_1 E_2)$. 
For the general single-species limit, the equal rate condition
$r_g=r_q$ requires the seven independent cross sections to satisfy 
Eq.\eqref{sigmagg}, i.e., 
\begin{equation}
\sigma_{gg} = \frac {3}{8} \left [\sigma_{qq} + \sigma_{q\bar q} 
+(N_f-1) (2\sigma_{qq^{\prime}} + \sigma_{q\bar q}^{q^{\prime}
    \bar q^{\prime}} ) \right ] +\frac {(4-3 N_f)}{32} \left (8
\sigma_{gq}+ 3  \sigma_{q\bar q}^{gg} \right ). 
\end{equation}
Applying the above relation, Eq.\eqref{etaNf} then reduces to the
following (without any $w_i$ terms): 
\begin{equation} 
\eta=\frac {1} {(x_g f_{gg} + 2 N_f x_q f_{gq} +  N_f x_g
  f_{gg}^{q\bar q})}~\frac{160\,T} {\Tilde{c}_{0}},
\label{etaNfLimit}
\end{equation}
where $\Tilde{c}_{0}$ has been defined in Eq.\eqref{c0}.

On the other hand, we expect the shear viscosity 
in this equal-rate limit to be the single-species result with an
effective cross section $\sigma_{\rm eff}$ that gives the same
collision rate per parton. From  $n\langle \sigma_{\rm eff} v_{\rm
  rel} \rangle=r_g$, we obtain the equivalent effective cross section as
\begin{equation} 
\sigma_{\rm eff}=\left (x_g f_{gg} + 2 N_f x_q f_{gq} +  N_f x_g
f_{gg}^{q\bar q} \right ) \sigma_0(v).
\end{equation}
The single-species result in Eq.\eqref{singleSpecies}  
for $\sigma=\sigma_{\rm eff}$ thus agrees with Eq.\eqref{etaNfLimit},
i.e., the Chapman-Enskog expression of shear viscosity in
Eq.\eqref{etaNf} satisfies this general single-species limit. 

\bibliography{refs}

\end{document}